\documentclass[dvips,a4paper]{article}

\usepackage{color}
\usepackage{amsmath,amssymb,amsfonts,epsfig}
\usepackage{verbatim}
\usepackage{multirow}
\usepackage{graphicx}
\usepackage{graphics}
\usepackage{url}
\usepackage{chngcntr}

\date{\today}

\newcommand{\ber}{\begin{eqnarray}}
\newcommand{\eer}{\end{eqnarray}}

\newcommand{\be}{\begin{equation}}
\newcommand{\ee}{\end{equation}}

\newcommand{\cs}{{\bf CS }}
\newcommand{\vv}{{\textbf v}}
\newcommand{\va}{{\textbf a}}
\newcommand{\vc}{{\textbf c}}
\newcommand{\vg}{{\textbf g}}
\newcommand{\vt}{{\textbf t}}
\newcommand{\vp}{{\textbf p}}
\newcommand{\bcs}{{\textbf {BCS} }}
\newcommand{\rRNA}{16S rRNA }

\newcommand{\nin}{\noindent}

\title{Bacterial Community Reconstruction Using A Single Sequencing Reaction}

\author{\renewcommand{\thefootnote}{\arabic{footnote}}
Amnon Amir\footnotemark[1]$\:^{,*}$ \quad
Or Zuk\footnotemark[2]$\:^{,*}$ }

\begin{document}

\maketitle
\footnotetext[1]{Department of Physics of Complex
Systems, The Weizmann Institute of Science, Rehovot, Israel}
\footnotetext[2]{Broad Institute of MIT and Harvard, Cambridge,
Ma, USA}
\let\thefootnote\relax\footnotetext{$^*$ \hspace{-0.08in} Equal contribution}

\abstract
Bacteria are the unseen majority on our planet, with millions of species and comprising most
of the living protoplasm. While current methods enable in-depth study of a small number of communities,
a simple tool for breadth studies of bacterial population composition in a large number of samples is lacking.
We propose a novel approach for reconstruction of the composition of an unknown mixture of bacteria using a
single Sanger-sequencing reaction of the mixture. This method is based on compressive sensing theory, which deals
with reconstruction of a sparse signal using a small number of measurements.
Utilizing the fact that in many cases each bacterial community is comprised of a small subset of the known bacterial species,
we show the feasibility of this approach for determining the composition of a bacterial mixture.
Using simulations, we show that sequencing a few hundred base-pairs of the 16S rRNA gene sequence may provide enough information
for reconstruction of mixtures containing tens of species, out of tens of thousands, even in the presence of realistic measurement noise.
Finally, we show initial promising results when applying our method for the reconstruction of a toy experimental mixture with five species.
Our approach may have a potential for a practical and efficient way for identifying bacterial species compositions in biological samples. \\
Availability: A MATLAB code is available at: \\
\url{http://www.broadinstitute.org/~orzuk/matlab/libs/BCS/matlab_BCS_utils.html}

\counterwithin{figure}{section}
\counterwithin{equation}{section}

\section{Introduction}
Microorganisms are present almost everywhere on earth. The population of bacteria found
in most natural environments consists of a large number of species, mutually
affecting each other, and creating complex ecological systems \cite{keller_tapping_2004}.
In the human body, the number of bacterial cells is over an order of magnitude
larger than the number of human cells \cite{Savage:01}, with typically several hundred species identified
in a given sample taken from humans (for example, over 400 species were characterized in the human gut
\cite{Eckburg:01}, while \cite{sears2005dynamic} estimates a higher number of 500-1000, and 500 to 600 species were
found in the oral cavity \cite{dewhirst2008human,paster_bacterial_2001}). Changes in the human bacterial community
composition are associated with physical condition, and may indicate \cite{Mager:01}
as well as cause or prevent various microbial diseases \cite{Guarner:01}.
In a broader aspect, studies of bacterial communities range from understanding the plant-microbe
interactions \cite{Singh:01}, to temporal and meteorological effects on the
composition of urban aerosols \cite{brodie_urban_2007}, and is a highly active field of research \cite{medini_microbiology_2008}.

Identification of the bacteria present in a given sample is not a simple task, and technical limitations
impede large scale quantitative surveys of bacterial community compositions. While conventional phenotypic methods (such as fatty acid profiles,
carbon source utilization and biochemical identification)
are relatively inexpensive, they often show an inaccurate and biased identification \cite{clarridge2004impact,woo2008then}.
Such methods were shown to
provide incorrect identification of many organisms (see \cite{bosshard2003ribosomal,tang2000identification}), and
additionally rely on the availability of pure culture of each bacteria present in the sample. Since the vast majority of bacterial
species are non-amenable to standard laboratory cultivation procedures \cite{Amann:01}, this leads to a highly biased identification.
Much attention has been therefore given to alternative, culture-independent methods.
The golden standard of microbial population analysis
has been direct Sanger sequencing of the ribosomal 16S subunit gene (\rRNA\!) \cite{hugenholtz_exploring_2002}.
Briefly, DNA from the bacterial population is extracted and PCR amplified using universal primers. The resulting \rRNA
gene is cloned and single colonies are sequenced using Sanger sequencing. This method offers high accuracy since
the whole \rRNA gene is sequenced and used for identification of each clone. However, the sensitivity of this
method is determined by the number of sequencing reactions, and therefore requires hundreds of sequences for each
sample analyzed. Due to the high cost and labor of such a process, using this method is limited to in-depth study
on a small number of samples (see for example \cite{Eckburg:01}).
A modification of this method for identification of small mixtures of bacteria using a single Sanger sequence has
been suggested \cite{kommedal_analysis_2008} and showed promising results when reconstructing mixtures of 2-3 bacteria
from a given database of $~260$ human pathogen sequences.

Recently, DNA microarray-based methods \cite{gentry_microarray_2006} and identification via next generation sequencing (reviewed in
\cite{hamady_microbial_2009}) have been used for bacterial community reconstruction. In microarray based methods,
such as the Affymetrix PhyloChip platform \cite{brodie_urban_2007}, the sample \rRNA is hybridized with
short probes aimed at identification of known microbes at various taxonomy levels. While being more sensitive and cheaper than
standard cloning and sequencing techniques, each bacterial mixture sample still needs to be hybridized against a microarray, thus
the cost of such methods limit their use for wide scale studies.
While most microarray-based methods require a design of arrays with a different probe specific to each one of the species to be detected,
a recent approach \cite{Dai:01} has proposed to use universal Compressed-Sensing based microarray design, in which one takes into account the
hybridization of a given probe to multiple (rather than single) different sequences.
Methods based on next generation sequencing obtain a very large number of reads of a short hyper-variable
region of the \rRNA \cite{armougom_use_2008,dethlefsen_pervasive_2008,hamady2008error}. By matching these sequence reads to
known \rRNA sequences, the bacterial composition is reconstructed. Usage of such methods, combined with DNA barcoding, enables
high throughput identification of bacterial communities, and can potentially detect species present at very low frequencies.
However, since such sequencing methods are limited to relatively short read lengths (typically a few dozens and at most a few hundred bases
in each sequence), the identification is non unique and limited in resolution, with reliable identification typically up to the genus level \cite{huse_exploring_2008}.
Improving the resolution depends on obtaining longer read lengths, which is currently technologically challenging.

\begin{figure}[!ht]
\begin{center}
\includegraphics[totalheight=0.6\textheight]{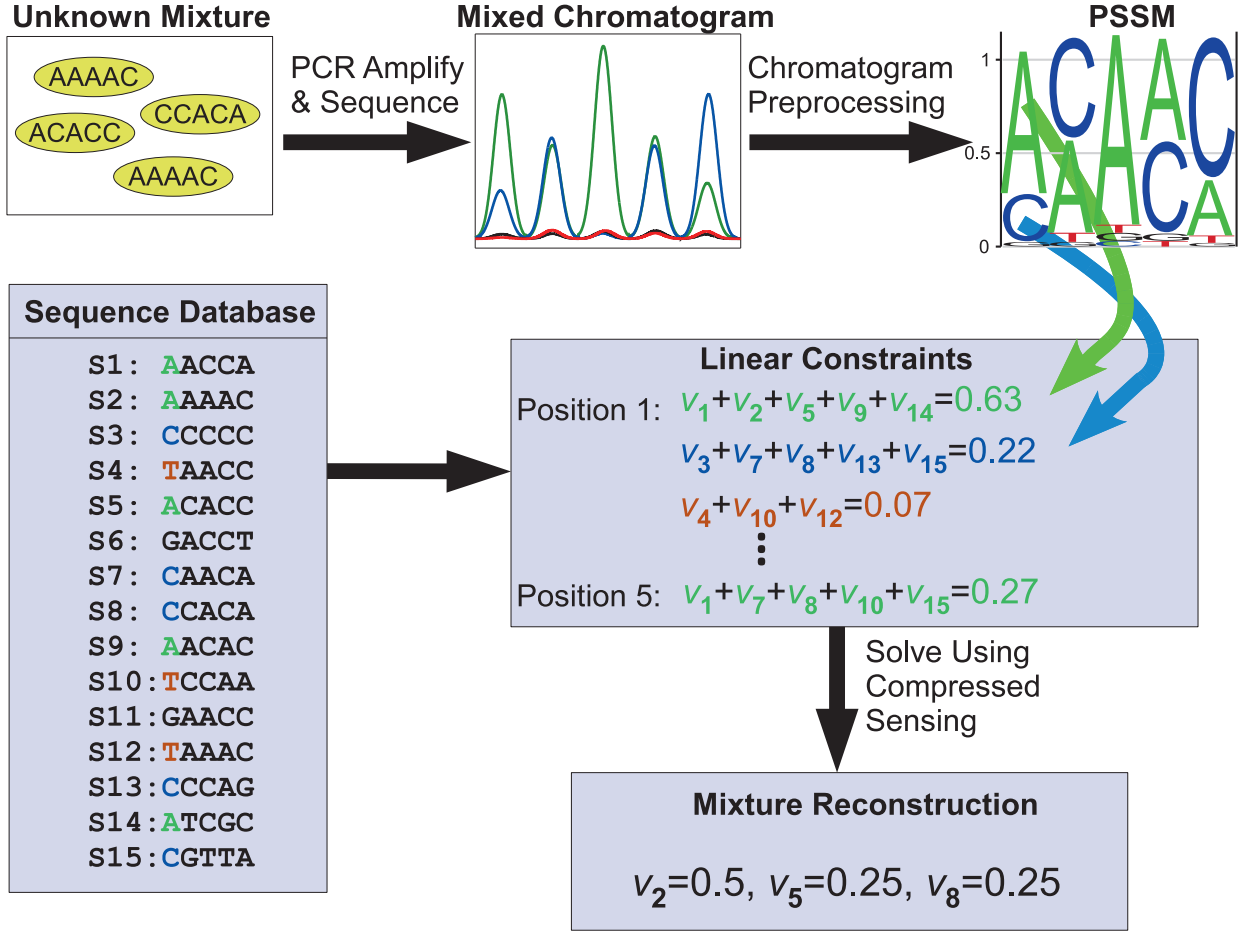}
\end{center}
\caption{
{\bf Schematics of the proposed \bcs reconstruction method.}
The \rRNA gene is PCR-amplified from the mixture and
then subjected to Sanger sequencing. The resulting chromatogram is preprocessed to create the Position Specific Score
Matrix (PSSM). For each sequence position, four linear mixture equations are derived from the \rRNA sequence database,
with $v_i$ denoting the frequency of sequence $i$ in the mixture, and the frequency sum taken from the experimental PSSM.
These linear constraints are used as input to the \cs algorithm, which returns the sparsest set of bacteria
recreating the observed PSSM.
}
\label{fig:MethodScheme}
\end{figure}

In this work we suggest a novel experimental and computational approach for sequencing-based profiling of bacterial communities
(see Figure \ref{fig:MethodScheme}). Our method relies on a single Sanger sequencing reaction for a bacterial mixture,
which results in a linear combination of the constituent sequences. Using this mixed chromatogram as linear constraints,
the sequences which constitute the original mixture are selected using a Compressed Sensing (\cs\!\!) framework.

Compressed Sensing (\cs\!) \cite{Candes:01,Donoho:01} is an emerging field of research, based on statistics and optimization
with a wide variety of applications.
The goal of \cs is recovery of a signal from a small number of measurements, by exploiting the
fact that most natural signals (e.g. natural images or human speech) are in fact sparse, or approximately sparse,
when represented at a certain appropriate basis.
Compressed Sensing designs sampling techniques that condense the information of a compressible signal into a small amount of data.
This offers the possibility of performing fewer measurements than were thought to be needed before, thus lowering costs and
simplifying data-acquisition methods for various types of signals in many distantly related fields such as magnetic resonance
imaging \cite{Lustig:01}, single pixel camera \cite{Duarte:01}, geophysics \cite{Lin:02} and astronomy \cite{Bobin:01}.

\cs was recently applied to various problems in computational biology, e.g. for pooling designs for re-sequencing experiments
\cite{erlich2009compressed,Shental:01}, and for drug-screenings \cite{kainkaryam2009pooling}.
Lately \cs was also used to design multiplexed DNA microarrays \cite{Dai:01}, where each spot is a combination of several
different probes. By specific selection and mixing of the probes, a \cs microarray was designed for detection of bacterial species in a community
using a small number of probe spots.

The classical problem in \cs is solving an under-determined linear system of equations
\be
\mathcal{A} \textbf{v} = \textbf{b}
\label{eq:cs_classic}
\ee
where $\textbf{v}=(v_1,...,v_N)$ is the vector of unknown variables, $\mathcal{A}$ is the {\it sensing} matrix, often called also the {\it mixing} matrix
and $\textbf{b}=(b_1,...,b_k)$ are the measured values of the $k$ equations. The number of variables $N$, is far greater than the
number of equations $k$. Without further information, $\textbf{v}$ cannot be reconstructed uniquely since the system is
under-determined. Here one uses an additional sparsity assumption on the solution - by assuming that we are interested only in
solution vectors $\textbf{v}$ with only at most $s$ non-zero entries, for some $s \ll N$. According to the \cs theory,
when the matrix $\mathcal{A}$ satisfy certain conditions, most notably the RIP condition \cite{candes2005stable,candes2005decoding},
one can find the sparsest solution uniquely by using only a logarithmic number of equations, $k = O(s \log(N/s))$, instead of a linear number (N)
needed for general solution of a linear system. Furthermore, efficient algorithms exist which make finding the solution
practical even for very large sensing matrices, with $N$ sizes up to tens of thousands of variables. Intuitively, the RIP condition states that the
columns of the matrix $A$ are not similar to each other, and in fact close to orthogonal. More details on the requirements from the matrix
$A$, the problem representation and algorithms for solutions of \cs problems are given in the Methods section, as well as in the
aforementioned references.

In this paper, we show an efficient application of pooled Sanger-sequencing for reconstruction of bacterial communities using \cs\!\!.
The sparseness assumption in our scenario is obtained by noting that although numerous species of bacteria have been characterized
and are present on earth, at a given sample typically only a small fraction of them are present at significant levels, while the proportion
of all other species is essentially zero. This assumption enables an accurate reconstruction of the non-zero frequencies, from a relatively
small amount of information, obtained by a single sequence of a certain gene such as the \rRNA gene. The proposed Bacterial Compressed
Sensing algorithm (\bcs\!\!) uses as inputs a database of known \rRNA sequences and the measured Sanger-sequence of the unknown
mixture, and returns the sparse set of bacteria present in the mixture and their predicted frequencies. The measured Sanger-sequence
we use as input comes as a chromatogram, representing a linear combination of DNA sequences coming from species in our mixture. This is different from
the commonly used chromatograms which typically represent a single sequence. We therefore developed a set of pre-processing steps adjusted
for such mixed chromatograms - the use of these steps, before applying the \cs optimization procedure, was crucial for the success of the
reconstruction. We show a successful reconstruction of simulated mixtures containing dozens of bacterial species out of a database of tens
of thousands, using realistic biological parameters. In addition, we demonstrate the applicability of our proposed method for a real sequencing
experiment using a toy mixture of five bacterial species.

\section{Methods}
\subsection{The Bacterial Community Reconstruction Problem}
In the Bacterial Community Reconstruction Problem we are given a bacterial mixture of unknown composition. In addition, we have
at hand a database of the orthologous genomic sequences for a specific known gene, which is assumed to be present in a large
number of bacterial species (in our case, the gene used was the \rRNA gene). Our purpose is to reconstruct the identity of species present in
the mixture, as well as their frequencies, where the assumption is that the sequences for the gene in all or the vast majority of
species present in the mixture are available in the database. The data used for the reconstruction problem is obtained from sequencing
the specific gene in the bacterial mixture. The input to the reconstruction algorithm is thus the measured
Sanger sequence of the gene in the mixture (see Figure \ref{fig:MethodScheme}). We used the \rRNA gene for reconstruction for
two reasons: first, its sequence is known for a very large number of bacteria, and databases with these sequences are available
\cite{desantis2006greengenes,maidak2001rdp}. Second, the gene contains both several highly conserved regions as well as variable
regions. The conserved regions can serve for universal amplification and sequencing of the gene using a single primer. The sequence
information from the variable regions provides the ability to distinguish between the different
species present in the mixture.
Since Sanger sequencing proceeds independently for each DNA strand present in the sample, the sequence chromatogram of the
mixture  corresponds to the linear combination of the constituent sequences, where the linear coefficients are proportional to the
abundance of each species in the mixture. Using this mixture sequence as linear constraints on the bacterial mixture composition,
we try to determine the composition of the mixture.

Let $N$ be the number of known bacterial species present in our database. Each bacterial population is characterized by
a vector $\vv=(v_1,...,v_N)$ of frequencies of the different species. Denote by $s$ the number of species present in the
sample $s=\|\vv\|_{\ell 0}$, where $\|.\|_{\ell 0}$ denotes the $\ell_0$ norm which simply counts the number of non-zero elements
of a vector $\|\vv\|_{\ell 0} = \sum_i 1_{\{v_i \neq 0\}}$. While the total number of known species $N$ is usually very large (in
our case on the order of tens to hundreds of thousands), a typical bacterial community consists of a small subset of the species,
and therefore in a given sample, $s \ll N$, and $\vv$ is a sparse vector. The database sequences are denoted by a matrix $S$, where $S_{ij}$
is the $j$'th nucleotide in the orthologous sequence of the $i$'th species ($i=1,..,N, j=1,..,k$). We represent the results of the mixture Sanger sequencing
of length $k$ as a $4 \times k$ Position-specific-Score-Matrix (PSSM)
\be
P =
\begin{pmatrix}
a_1 & a_2 & \cdots & a_k \\
c_1 & c_2 & \cdots & c_k \\
g_1 & g_2 & \cdots & g_k \\
t_1 & t_2 & \cdots & t_k \\
\end{pmatrix} =
\begin{pmatrix}
\va \\
\vc \\
\vg \\
\vt \\
\end{pmatrix}
\label{eq:PSSM}
\ee
where $\vp_j=(a_j, c_j, g_j, t_j)^t$ is a column vector representing the observed frequencies at sequence position $j$ of
the four nucleotides, with $a_j,c_j,g_j,t_j \geq 0$.

Each position in the mixed sequence gives information about the bacterial composition
of the mixture. For example, if at a certain position $j$, the frequency $a_j$ of '$A$' in the mixed sequence is $0$,
and assuming no measurement noise is present, it follows that all bacteria which have '$A$' at the $j$'th position of their orthologous gene
are not present in the mixture. More generally, the frequency of each nucleotide at a given position $j$ gives a linear constraint on the mixture:
\be
\sum_{i=1}^N v_i 1_{\{ S_{ij} =  '\!A' \}} = a_j
\ee
and similarly for $'\!C','\!G','\!T'$.

Define the $k \times N$ mixture matrix $A$ for the nucleotide '${A}$':
\be
A_{ij}=\left\{ \begin{array}{ll}
1 & S_{ij} = '\!\!{A}' \\
0 & \mbox{otherwise} \end{array} \right.
\label{eq:MixMatrix}
\ee
and similarly for the nucleotides $'\!{C}','\!{G}','\!{T}'$.
The constraints given by the sequencing reaction can therefore be expressed as:
\be
A\textbf{v}=\textbf{a}, C\textbf{v}=\textbf{c}, G\textbf{v}=\textbf{g}, T\textbf{v}=\textbf{t} \label{eq:LinConstraints}
\ee

Hence, from a single sequencing reaction of length $k$ we derive a set of $4k$ linear equations. Since a typical
sequencing reaction is of length $k \sim 500\!-\! 1000$, the total number of equations is $4k \sim 2000 \! - \! 4000$. This is
still considerably smaller than the number of free variables $N$ which is on the order of tens to hundreds of thousands.
Hence, this is an underdetermined linear system, and without further assumptions we are not guaranteed a single solution.
The crucial assumption we make in order to cope with the insufficiency of information is the sparsity of the vector $\textbf{v}$,
which reflects the fact that only a small number of species are present in the mixture. This assumption enables us to formulate
and solve the reconstruction as a \cs problem, as described in the next section.

\subsection{Mapping the Problem to Compressed Sensing - the \bcs Algorithm}
We seek a sparse solution for the set of equations (\ref{eq:LinConstraints}). Such a solution provides a small set of species
which is consistent with the measured data. In the \cs paradigm, one can formulate the quest for
the sparsest solution satisfying the given set of linear constraints as follows:
\be
\textbf{v}^*= \underset{\textbf{v}}{argmin}\|\textbf{v}\|_{\ell 0} \quad s.t.
\quad A\textbf{v}=\textbf{a}, C\textbf{v}=\textbf{c}, G\textbf{v}=\textbf{g}, T\textbf{v}=\textbf{t}
\label{eq:L0Min}
\ee
The formulation above can be easily seen as a \cs problem - one can in principle construct a big 'sensing' matrix $\mathcal{A}_{4k\times N}$
as a simple concatenation of the matrices A,C,G,T, and similarly a measurements vector $\textbf{b}$ as a concatenation of
the measurements $\textbf{a}, \textbf{c}, \textbf{g}$ and $\textbf{t}$, and with these notations
problem (\ref{eq:L0Min}) is in fact reduced to the classic \cs problem \cite{Candes:01} of finding the sparsest solution for
eq. (\ref{eq:cs_classic}). From a practical point of view, problem (\ref{eq:L0Min}) is known to be in general
NP-Hard \cite{candes2005error}, and hence finding the optimal solution is not computationally feasible. A main cause for the hardness of the
problem is the non-convex $\ell 0$ constraint. However, a remarkable breakthrough of the \cs theory shows that under certain conditions on
the mixture matrix and the number of measurements (see below), the sparse solution can be recovered uniquely by solving a convex relaxation
of the problem, which is obtained by replacing the $\ell 0$ norm with the ''closest'' convex $\ell p$ norm, namely the $\ell 1$ norm,
leading to the following minimization problem \cite{candes2006near,donoho2006most,tropp_just_2006}:
\be
\textbf{v}^*=\underset{\textbf{v}}{argmin}\|\textbf{v}\|_{\ell 1} = \underset{\textbf{v}}{argmin}\sum_{i=1}^N |v_i| \quad s.t.
\quad A\textbf{v}=\textbf{a}, C\textbf{v}=\textbf{c}, G\textbf{v}=\textbf{g}, T\textbf{v}=\textbf{t}
\label{eq:L1Min}
\ee
which is a convex optimization problem whose solution can be obtained in polynomial time. The above formulation requires our
measurements to be precisely equal to their expected value based on the species frequency and the linearity assumption for the
measured chromatogram. This description ignores the effects of noise, which is typically encountered in practice, on the reconstruction.
Clearly, measurements of the signal mixtures suffer from various types of noise and biases. Fortunately, the \cs paradigm is known
to be robust to measurement noise \cite{candes2005stable,candes2007dantzig}. One can cope with noise by enabling a
trade-off between sparseness and accuracy in the reconstruction merit function, which in our case is formulated as:
\be
\textbf{v}^* = \underset{\textbf{v}}{argmin} \:\: \frac{1}{2} \big( \|\va-A\textbf{v}\|^2_{\ell 2} + \|\vc-C\textbf{v}\|^2_{\ell 2} +
\|\vg-G\textbf{v}\|^2_{\ell 2} + \|\vt-T\textbf{v}\|^2_{\ell 2} \big) + \tau \|\textbf{v}\|_{\ell 1}
\label{eq:GPSR}
\ee

This problem represents a more general form of eq. (\ref{eq:L1Min}), and accounts for noise in the measurement process. This is utilized by insertion of an $\ell 2$
quadratic error term. Importantly, even with the addition of the $\ell 2$ term the problem (\ref{eq:GPSR}) is still a convex optimization problem.
The parameter $\tau$ determines the relative weight of the error term vs. the sparsity promoting term. Increasing $\tau$ leads to a sparser solution,
at the price of a worse fit for the measurement equations, whereas decreasing $\tau$ leads to a better fit to the equations, while possibly requiring
more non-zero elements in the solution (for low enough values of $\tau$ we can fit the equations in (\ref{eq:L1Min}) precisely, and the $\ell 2$ error term vanishes, thus
the problem is reduced back to eq. (\ref{eq:L1Min})).
Many algorithms which enable an efficient solution of problem (\ref{eq:GPSR}) are available, and
we have chosen the widely used GPSR algorithm described in \cite{GPSR:01}.
The error tolerance parameter was set to $\tau=10$ for the simulated mixture reconstruction, and $\tau=100$ for the reconstruction of the experimental mixture.
These values achieved a rather sparse solution in most cases (a few species reconstructed with frequencies above zero), while still giving a good sensitivity.
The performance of the algorithm was quite robust to the specific value of $\tau$ used, and therefore further optimization of the results by fine tuning $\tau$ was not
followed in this study. Accuracy of solution was evaluated using two measures, Root-Mean-Squared-Error (RMSE) and recall/precision. For the latter measure
we have set a minimal frequency threshold for inclusion in both the true and the reconstructed solutions. More detailed are provided in the Results section.

In classical \cs one designs the mixing matrix to have certain desirable properties in order to enable unique reconstruction.
The most well-known condition for successful reconstruction is the 'Robust Isometry Property' (RIP), also known as the
'Uniform Uncertainty Principle' (UUP) \cite{candes2005stable,candes2005decoding}. Briefly, RIP for a matrix means that any
subset of $2s$ columns of the matrix $\mathcal{A}$ is 'almost orthogonal' (although since $k < N$ the columns cannot be
perfectly orthogonal). This property makes the matrix $\mathcal{A}$ 'invertible' for sparse vectors $v$ with sparsity $s$. Furthermore, it is
known that with very high probability a unique and accurate reconstruction is achievable with as few as $O(s \log (N/s))$
measurements, compared to $O(N)$ measurements required for finding a solution for a general linear system without a sparsity
assumption \cite{Candes:01,Donoho:01}. In our case the number of measurements $4k$ is on order of a few thousands, thus
these results suggest that accurate reconstruction is possible, at least when the sparsity $s$ in the range of a few dozens. One important
difference between classical \cs and our problem is the structure of the mixing matrix $\mathcal{A}$. In our application
the mixing matrix represents the sequence database $S$ which is pre-determined. Moreover, for species which are close in the
bacterial phylogeny, the gene sequences in the database exhibit high sequence similarity and therefore the corresponding
columns in the sensing matrix are far from orthogonal. It is thus far from clear in advance that the given mixing matrix posses the
desired properties to enable successful reconstruction. This point is addressed in the Results section.

\subsection{Ribosomal DNA Database}
\rRNA gene sequences were obtained from grenegenes (\url{greengenes.lbl.gov}) using database version 06-2007 \cite{desantis2006greengenes},
which contains approximately $136000$ chimera checked full length sequences. Sequences were reverse complemented and aligned
with primer 1510R \cite{gao_molecular_2007}, resulting in approximately $42000$ sequences matching the primer sequence (with up
to 6 mismatches with the primer). Out of this set, sequences with up to 2 base-pair difference with another sequence in the database
were removed, resulting in $N=18747$ unique sequences which were used in this study. This last step was used in order to reduce the size of the
input to the GPSR algorithm, thus enabling solution of the \cs problem using a standard PC.

The sequence of \textit{Enterococcus faecalis} (ATCC \# 19433) was manually added to the list of unique sequences, as it did not appear in the database
(closest neighbor in the database has 32 different positions), and is used in the experimental mixture.

\subsection {Experimental Mixture Reconstruction}
\subsubsection{Sample Preparation}
Strains used for the experimental reconstruction were:
\textit{Escherichia coli W3110, Vibrio fischeri, Staphylococcus epidermidis} (ATCC \# 12228),
\textit{Enterococcus faecalis} (ATCC \# 19433) and \textit{Photobacterium leiognathi}.
The \rRNA gene was obtained from each bacterial strain by boiling for one minute
followed by 40 cycles of PCR amplification. Primers used for the PCR were the universal primers 8F and
1510R \cite{gao_molecular_2007}, amplifying positions 8-1513 of the \textit{E. coli} \rRNA\!:

\hspace{0.022cm} \textbf{8F}: 5'-AGAGTTTGATYMTGGCTCAG \\
\textbf{1510R}: 5'-TACGGYTACCTTGTTACGACTT

For mixture preparation and sequencing, equal amounts of DNA from each bacterial \rRNA gene were mixed together,
and then sequenced using an ABI3730 DNA Analyzer (Applied Biosystems, USA) using the 1510R primer.

\subsubsection{Preprocessing Steps}
The input to the \bcs algorithm is a $4 \times k$ PSSM $(\va, \vc, \vg, \vt)^t$ of the mixture. However, obtaining this PSSM from an experimental
mixture is not trivial. The output of a Sanger-sequencing reaction is a chromatogram, which describes the fluorescence of the four
terminal nucleotides as a function of sequence position. In classical single-species sequencing, each peak in the chromatogram corresponds to
a single nucleotide in the sequence. Identifying the peaks becomes more complicated when sequencing a mixture of different sequences.
It has been previously shown  (see e.g. \cite{bowling_neighboring_1991,nickerson1997polyphred}) that
chromatogram peak height and position depend on the local sequence of nucleotides preceding a given nucleotide.
Therefore, when performing Sanger sequencing of a mixture of multiple DNA sequences, the peaks
of the constituent sequences may lose their coherence, making it nearly impossible to determine where the chromatogram peaks are located.
We therefore opted for a slightly different approach for preprocessing of the chromatogram, which does not depend on identifying the
peak for each nucleotide. Rather, the chromatogram is binned into constant sized bins, and the total intensity of each of the four
nucleotides in each bin is used to construct the PSSM used as input to the \bcs (see Figure \ref{fig:PreprocessingScheme}.A).
A similar process is applied to each sequence in the 16S rRNA database. In order to correct for local-sequence effects, statistics were collected for local-sequence
dependence of peak height and position. Similar statistics
are used to obtain quality scores for single-sequence chromatogram base-calling in the Phred algorithm \cite{ewing1998base2,ewing1998base1}.
By utilizing these statistics, we predict the
chromatogram for each sequence in the database, which is then binned and results in a PSSM for the single sequence. This database of predicted PSSMs
is then used to construct the mixing matrices $A, C, G, T$ participating in the \bcs problem representation in eq. (\ref{eq:GPSR}) (see Figure \ref{fig:PreprocessingScheme}.B). The details of
the preprocessing steps for the database sequences and the measured chromatogram are given in the appendix.

\section{Results}
\subsection{Simulation Results}
In order to asses the performance of the proposed \bcs reconstruction algorithm, random subsets of species from the greengene database
\cite{desantis2006greengenes} were selected. Within these subsets, the relative frequencies of each species were drawn at random from
a uniform frequency distribution (normalized to sum to one), and a mixed sequence PSSM was calculated (results for a different, power-law
frequency distribution, are shown later).
This PSSM was then used as the input for the \bcs algorithm, which returned the frequencies
of database sequences predicted to participate in the mixture (see Methods section and Figure \ref{fig:MethodScheme}).

\begin{figure}[!ht]
\begin{center}
\includegraphics[totalheight=0.4\textheight]{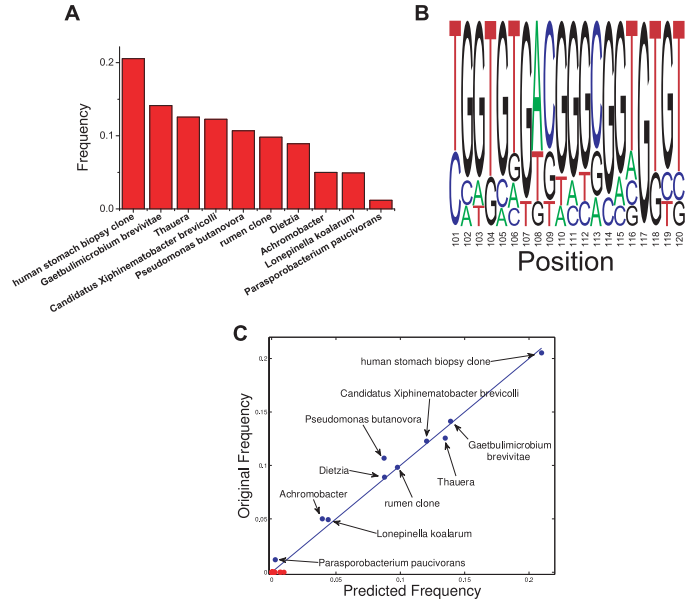}
\end{center}
\caption{
\textbf{Sample reconstruction of a simulated mixture.}
\textbf{A.} Frequencies and species for a simulated random mixture of $s=10$ sequences. Species were randomly selected
from the \rRNA database, with frequencies generated from a uniform distribution.
\textbf{B.} A 20 nucleotide sample region of the PSSM for the mixture in (A).
\textbf{C.} True vs. predicted frequencies for a sample \bcs reconstruction for the mixture in (A) using $k=500$ bases of the
simulated mixture. Red circles denote species returned by the \bcs algorithm which are not present in the original mixture.}
\label{fig:SampleReconstruction}
\end{figure}

A sample of a random mixture of 10 sequences, and a part of the corresponding mixed sequence PSSM, are shown in
Figure \ref{fig:SampleReconstruction}.A,B respectively. Results of the \bcs reconstruction using a 500 bp long sequence
are shown in Figure \ref{fig:SampleReconstruction}.C. The \bcs algorithm successfully identified all of the species present
in the original mixture, as well as several false positives (species not present in the original mixture). The largest false positive
frequency was $0.01$, with a total fraction of $0.04$ false positives. In order to quantify the \bcs algorithm's performance,
we used two main measures: RMSE and recall/precision. RMSE is the Root-Mean Squared-Error between the original
mixture vector and the reconstructed vector, defined as
$RMSE = \|\textbf{v}-\textbf{v}^*\|_{\ell 2} = \Big(\sum_{i=1}^N (v_i-v^*_i)^2\Big)^{1/2}$. This measure accounts both
for the presence or absence of species in the mixture, as well as their frequencies.
In the example shown in Figure \ref{fig:SampleReconstruction} the RMSE score of the reconstruction was $0.03$.
As another measure, we have recorded the {\it recall}, defined as the fraction of species present in the original vector $\textbf{v}$
which were also present in the reconstructed vector $\textbf{v}^*$ (this is also known as sensitivity), and the {\it precision},
defined as the fraction of species present in the reconstructed vector $\textbf{v}^*$ which were also present in the original
mixture vector $\textbf{v}$. Since the predicted frequency is a continuous variable, whereas the recall/precision relies on a binary
categorization, a minimal threshold for calling a species present in the reconstructed mixture was used before calculating the
recall/precision scores.

\subsubsection{Coherence of Database Sequences}
\begin{figure}[!ht]
\begin{center}
\includegraphics[totalheight=0.4\textheight]{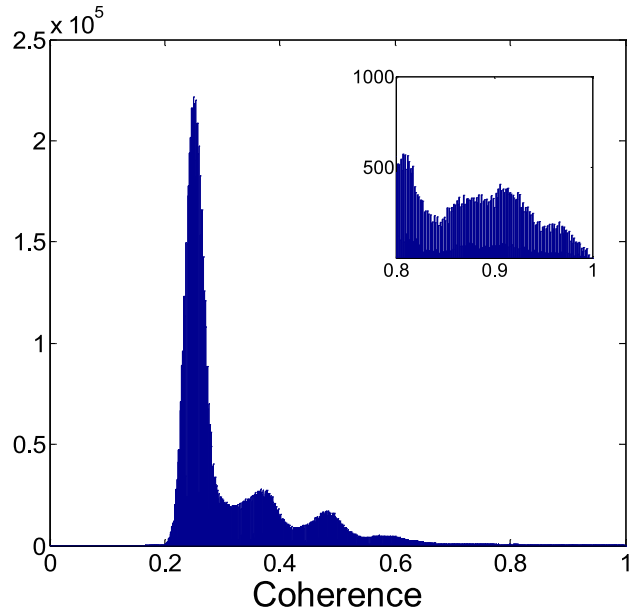}
\end{center}
\caption{
\textbf{Coherence distribution of the \rRNA sequences. }
Coherence (inner product) of $10^7$ \rRNA vector pairs chosen randomly from the sequence database ($\sim 5.7\%$ of all possible pairs).
As each column of the mixture matrix is a binary vector with $1/4$ of the coordinates being one, the dot product between two randomly
generated vectors is expected to be $\sim\!0.25$. While most \rRNA database pairs exhibit a coherence around $0.25$,
many pairs exhibit significantly higher correlations, with a few $(\sim\!0.5 \%$) even exceeding $0.8$ (see inset).}
\label{fig:Coherence}
\end{figure}

As explained in the Methods section, for successful reconstruction using a small number of measurements,
the columns of the mixing matrix need to be incoherent, i.e. close to orthogonal, in accordance with the
RIP condition \cite{Candes:02}.
In our case this cannot be achieved, as we were given the sequences determining the mixing matrix and cannot control
them. Even though the sequences are orthologous and thus quite similar, insertions and deletions came to our aid,
as they bring similar sequences to being out of phase (for example, even a deletion of a single base from a sequence,
reduces its correlation with a copy of itself from one to a number typically much lower).
It has been previously shown \cite{ben-haim_near-oracle_2009,tropp_just_2006} that a computationally
feasible method for assessing the information content of the mixing
matrix is the mutual coherence, defined as the maximal coherence (inner product) between two columns of the mixing matrix.
The distribution of coherence values for random pairs of database species is shown
in Figure \ref{fig:Coherence}. While most correlations are centered around $0.25$, there exists a small fraction of highly
correlated sequences, with $0.005$ of the sequence pairs showing a correlation above $0.8$, and a maximal correlation value
of $0.998$. This high mutual coherence value places a limit on the reconstruction performance in the worst case, when such a
sequence is present in the mixture. Since the database contains another highly similar sequence,
distinguishing between these two is very difficult, and therefore the \cs reconstruction cannot guarantee complete
accuracy. However, given that such similar sequences are typically of closely related species (thus not being able to distinguish
between them may be considered acceptable), and since most of the sequences show near random coherence, the
reconstruction in most of the cases may still require only a small number of measurements (which translates into a small number
of nucleotides read in the sequencing).

\subsubsection{Effect of Sequence Length}
\begin{figure}[!ht]
\begin{center}
\includegraphics[totalheight=0.2\textheight]{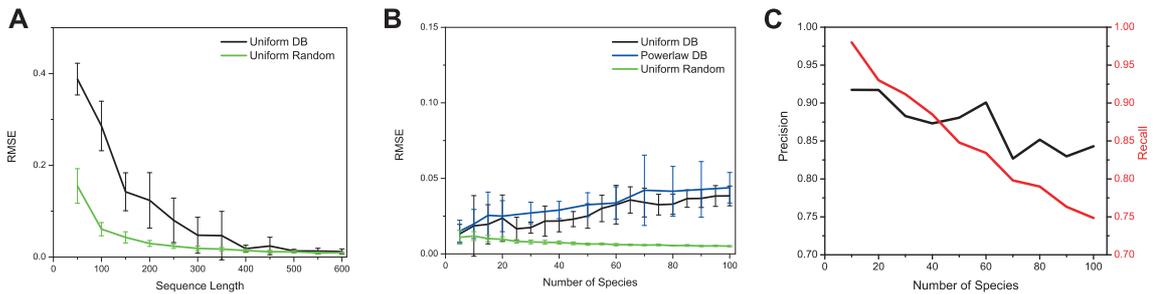}
\end{center}
\caption{
\textbf{Reconstruction of simulated mixtures.}
\textbf{A.} Effect of sequence length on reconstruction performance. RMSE between the original and reconstructed frequency
vectors for uniformly distributed random mixtures of $s=10$ species from the \rRNA database (black) or randomly generated sequences (green).
Error bars denote the standard deviation derived from $20$ simulations.
\textbf{B.} Dependence of reconstruction performance on number of species in the mixture. Simulation is similar to (A) but
using a fixed sequence length $(k = 500$) and varying the number of species in the mixture. Blue line shows reconstruction performance
on a mixture with power-law distributed species frequencies ($v_i \sim i^{-1}$).
\textbf{C.} Recall (fraction of sequences in the mixtures identified, shown in red) and precision (fraction of incorrect
sequences identified, shown in black) of the \bcs reconstruction of \rRNA the uniformly distributed database mixtures shown as black line in (B). The minimal
reconstructed frequency for a species to be declared as present in the mixture was set to $0.25\%$.}
\label{fig:SeqNum}
\end{figure}

To determine the typical sequence length required for reconstruction, we tested the \bcs algorithm performance using different sequence lengths.
In Figure \ref{fig:SeqNum}.A (black line) we plot the RMSE of reconstruction for random mixtures of $10$ species. To enable faster
running times, each simulation used a random subset of $N=5000$ sequences out of the sequence database for mixture generation
and reconstruction. It is shown in Figure \ref{fig:SeqNum}.A that using longer sequence lengths results in a larger number of linear constraints
and therefore higher accuracy, with $\sim\! 300$ nucleotides sufficing for accurate reconstruction of a mixture
of 10 sequences. The large standard deviation is due to a small probability of selection of a similar but incorrect sequence in the
reconstruction, which leads to a high RMSE.
Due to a cumulative drift in the chromatogram peak position prediction, typical usable experimental chromatogram lengths are in
the order of $k \sim\! 500$ bases rather than the $\sim 1000$ bases usually obtained when sequencing a single species
(see Methods section for details).

In order to asses the effect of the dependence between the database sequences (which leads to high coherence of the mixing matrix
columns) on the performance of the \bcs algorithm, a similar mixture simulation was performed using a database of random nucleotide sequences
(i.e. each sequence was composed of i.i.d. nucleotides with $0.25$ probability for 'A','C','G' or 'T'). Since sequences were independently
drawn, the pairwise correlation values are centered around $0.25$, with maximal coherence value less than $0.4$ over all pairs we have
simulated (data not shown). Using a mixing matrix derived from these random sequences, the \bcs reconstruction algorithm showed better
performance (green line in Figure \ref{fig:SeqNum}.A), with $\sim\! 100$ nucleotides sufficing for a similar RMSE as that obtained for the \rRNA
database using $300$ nucleotides.

\subsubsection{Effect of Number of Species}
For a fixed value of $k=500$ nucleotides per sequencing run, the effect of  the number of species present in the mixture on
reconstruction performance is shown in Figure \ref{fig:SeqNum}.B,C. Even on a mixture
of $100$ species, the reconstruction showed an average RMSE less than $0.04$, with the highest false positive reconstructed frequency (i.e. frequency
for species not present in the original mixture) being less than $0.01$. Using a minimal
frequency threshold of $0.0025$ for calling a species present in the reconstruction, the \bcs algorithm shows an average recall of
$0.75$ and a precision of $0.85$.
Therefore, while the sequence database did not perform as well as random sequences, the \rRNA sequences exhibit
enough variation to enable a successful reconstruction of mixtures of tens of species with a small percent of errors.

\begin{figure}[!ht]
\begin{center}
\includegraphics[totalheight=0.2\textheight]{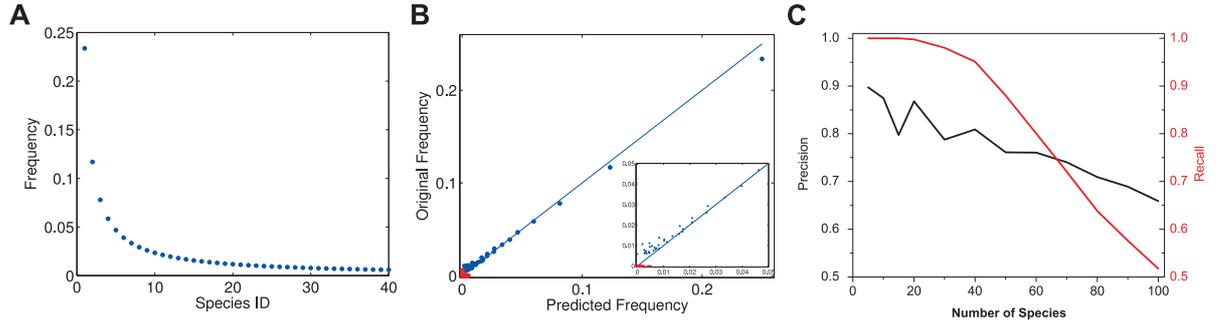}
\end{center}
\caption{
\textbf{Sample reconstruction of a power-law mixture.}
\textbf{A.} Sorted frequency distribution of 40 random species following a power-law distribution with frequencies $v_i \sim i^{-1}, i=1,..,40$.
\textbf{B.} True vs. predicted frequencies for a sample \bcs reconstruction for the mixture in (A) using $k=500$ bases of the
simulated mixture. Red circles denote species returned by the \bcs algorithm which are not present in the original mixture.
\textbf{C.} Average precision (black) and recall (red) for the reconstruction of simulated mixtures with power-law distributed frequencies as in (A).
The minimal reconstructed frequency for a species to be declared as present in the mixture was set to $0.17\%$.
}
\label{fig:PowerlawReconstruction}
\end{figure}

The frequencies of species in a biologically relevant mixture need not be uniformly distributed. For example, the frequency of species found on the
human skin \cite{gao_molecular_2007} were shown to resemble a power-law distribution. We therefore tested the performance of the \bcs reconstruction
on a similar power-law distribution of species frequencies with with $v_i \sim i^{-1}$. Performance on such a power-law mixture is similar
to the uniformly distibuted mixture (blue and green lines in Figure \ref{fig:SeqNum}.B respectively) in terms of the RMSE. A sample power-law mixture and
reconstruction are shown in Figure \ref{fig:PowerlawReconstruction}4.A,B. The recall/precision of the \bcs algorithm on such mixtures (Figure \ref{fig:PowerlawReconstruction}.C) is similar to
the uniform distribution for mixtures containing up to 50 species, with degrading performance on larger mixtures, due to the long tail of low frequency species.

\subsubsection{Effect of noise on \bcs solution}
\begin{figure}[!ht]
\begin{center}
\includegraphics[totalheight=0.25\textheight]{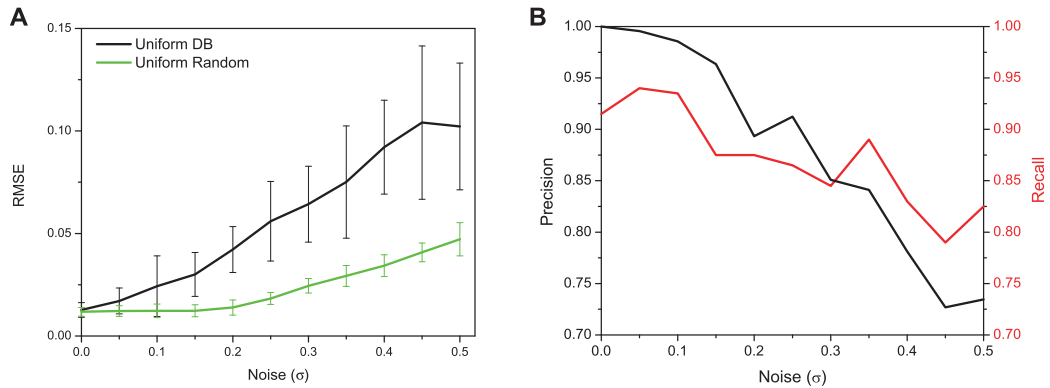}
\end{center}
\caption{
 \textbf{Effect of noise on reconstruction. }
\textbf{A.} Reconstruction RMSE of mixtures of $s=10$ sequences of length $k=500$ from the \rRNA sequence
database (black) or random sequences (green), with added normally distributed noise to the chromatogram.
\textbf{B.} Recall (red) and precision (black) of the \rRNA database mixture reconstruction shown in (A).}
\label{fig:SeqNoise}
\end{figure}

Experimental Sanger sequencing chromatograms contain inherent noise, and we cannot expect to obtain exact measurements in practice.
We therefore turned to study the effect of noise on the accuracy of the \bcs reconstruction algorithm. Measurement noise was modeled
as additive i.i.d. Gaussian noise $z_{ij}\sim N(0,\sigma^2)$ applied to each PSSM element in eq. (\ref{eq:PSSM}). Noise is compensated for
by the insertion of the $\ell 2$ norm into the minimization problem (see eq. (\ref{eq:GPSR})), where the factor $\tau$ determines the
balance between sparsity and error-tolerance of the solution. The effect of added random i.i.d.
Gaussian noise to each nucleotide measurement is shown in Figure \ref{fig:SeqNoise}. The reconstruction performance slowly degrades
with added noise both for the real \rRNA and the random sequence database.

\begin{figure}[!ht]
\begin{center}
\includegraphics[totalheight=0.2\textheight]{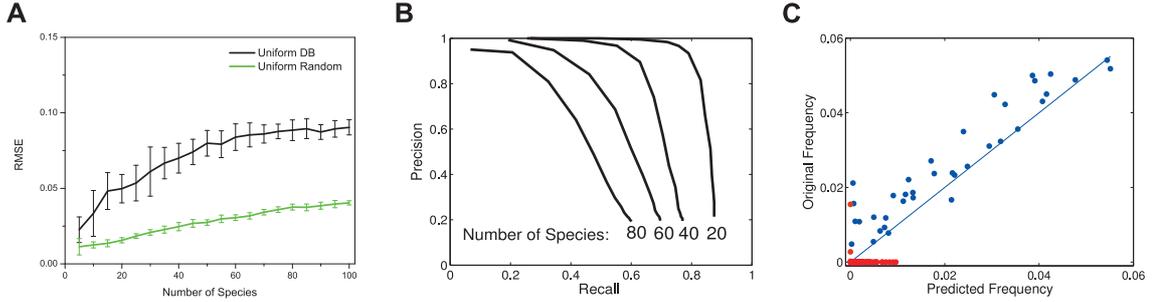}
\end{center}
\caption{
\textbf{Reconstruction with experimental noise level.}
\textbf{A.} Reconstruction RMSE as a function of number of species present in the mixture. Frequencies were sampled from a
uniform distribution. Noise is set to $\sigma=0.15$. Sequence length is set to $k=500$.
Black and green lines represent \rRNA and random sequences respectively.
\textbf{B.} Recall vs. precision curves for different number of \rRNA sequences as in (A) obtained by varying the minimal
inclusion frequency threshold.
\textbf{C.} Sample reconstruction of $s=40$ \rRNA sequences from (A).}
\label{fig:SampReconstruction}
\end{figure}

Using a noise standard deviation of $\sigma=0.15$ (which is the approximate experimental noise level - see later) and
sequencing  $500$ nucleotides, the reconstruction performance as a function of the number of species in the
mixture is shown in Figure \ref{fig:SampReconstruction}. Under this noise level, the \bcs algorithm reconstructed a mixture of
40 sequences with an average RMSE of $0.07$ (Figure \ref{fig:SampReconstruction}.B), compared to $\sim\!0.02$ when no
noise is present (Figure \ref{fig:SeqNum}.B). By using a minimal frequency threshold of $0.006$ for the predicted mixture,
\bcs showed a recall (sensitivity) of $\sim\! 0.7$, with a precision of $\sim\! 0.7$ (see Figure \ref{fig:SampReconstruction}.B),
attained under realistic noise levels. To conclude, we have observed that the addition of noise leads to a graceful degradation in the reconstruction
performance, and one can still achieve accurate reconstruction with realistic noise levels.

\subsection{Reconstruction of an Experimental Mixture}
We have applied our method to the problem of reconstruction of experimentally measured mixture chromatograms. This introduces a new problem of
interpreting non-ideal data, mainly that Sanger sequencing chromatograms exhibit a large variability in the peak heights and positions
(see Figure \ref{fig:PosHeightDist}). It has been previously shown that a large part of this variability stems from
local sequence effects on the polymerase activity \cite{lipshutz_dna_1994}.
While for standard sequencing applications this does not pose a problem, as only the qualitative information is required (which peak shows the maxima),
in our application this variability may prohibit the reconstruction. Since the measured chromatogram is a linear combination of the chromatograms
of the constituting bacterial strains, variability in peak position may cause a loss of phase between the chromatogram peaks, leading to possibly
overlapping peaks corresponding to different nucleotides. In order to overcome this problem, we utilize the fact that both peak position and height
are local sequence dependent in order to accurately predict the chromatograms of the sequences present in the 16S rRNA database.
The \cs problem is then stated in terms of reconstruction of the measured chromatogram using a sparse subset of predicted
chromatograms for the 16S rRNA database. This is achieved by binning both the predicted chromatograms and the measured mixture chromatogram
into constant sized bins, and applying the \bcs algorithm on these bins (see Methods section and Figure \ref{fig:PreprocessingScheme}).

A single-sequence chromatogram (measured with an ABI3730 sequencer) shows a standard deviation of approx. $0.26$ and $1.3$ for peak heights
and peak-peak distances respectively (red bars in Figure \ref{fig:SampleChromatogram}.B,C). In order to predict the effect of
local sequence of peak height and position, we considered the preceding 5 nucleotides for each sequence position (see
Methods). Statistics for the effect of each 6-mer were collected from ~1000 sequencing chromatograms performed on the ABI3730 sequencer.
Using this data, we can predict the peak positions and heights for a given sequence (see Figure \ref{fig:SampleChromatogram}.A).
The distribution of relative peak height and position errors following local sequence correction is shown in blue bars in
Figure \ref{fig:SampleChromatogram}.B,C, and the standard deviation is reduced to $0.15$ and $0.75$ for peak height and position respectively.

\begin{figure}[!ht]
\begin{center}
\includegraphics[totalheight=0.6\textheight]{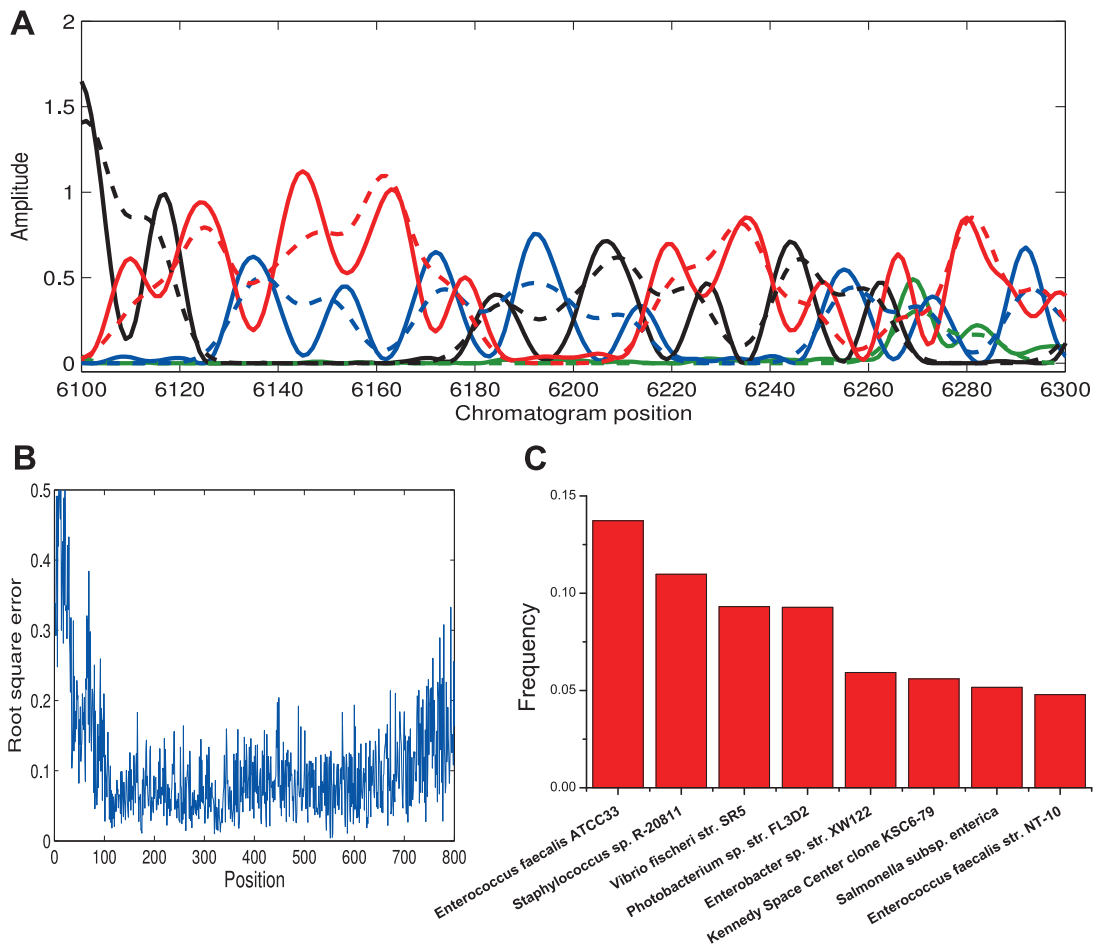}
\end{center}
\caption{
\textbf{Reconstruction of an experimental mixture.}
\textbf{A.} Sample region of the mixed chromatogram (solid lines). \rRNA from five bacteria was extracted and mixed at
equal proportion. dotted lines show the local-sequence corrected prediction of the chromatogram using the known mixture sequences.
\textbf{B.} Root square distance between the predicted and measured chromatograms shown in (A) as a function of the bin position, representing nucleotide
position in the sequence. Sequence positions in the range 100-700 achieved a low prediction error.
\textbf{C.} Reconstruction results using the \bcs algorithm. Runtime was $\sim\! 1000$ seconds on a standard PC.
Shown are the 8 most frequent species.
Original strains were : \textit{Escherichia coli, Vibrio fischeri, Staphylococcus epidermidis, Enterococcus faecalis} and
\textit{Photobacterium leiognathi}, with frequency $0.2$ each.
}
\label{fig:MixChromatogram}
\end{figure}

Using these chromatogram predictions, we tested the feasibility of the \bcs algorithm on experimental data by reconstructing a simple bacterial population
using a single Sanger sequencing chromatogram. We used a mixture of five different bacteria: (\textit{Escherichia coli W3110, Vibrio fischeri,
Staphylococcus epidermidis, Enterococcus faecalis} and \textit{Photobacterium leiognathi}).
DNA was extracted from each bacteria, and the
\rRNA gene was PCR amplified using universal primers 8F and 1510R (see Methods). The resulting \rRNA gene was mixed in
equal proportions and the mixture was sequenced using the universal 1510R primer. Data from the resulting chromatogram was used as input to the \bcs
algorithm following preprocessing steps described in the Methods section. A sample of the measured chromatogram is
shown in Figure \ref{fig:MixChromatogram}.A (solid lines).
The \bcs algorithm relies on accurate prediction of the chromatograms of each known database 16S rRNA sequence.
In order to asses the accuracy of these predictions, Figure \ref{fig:MixChromatogram}.A shows a part of the predicted chromatogram of the mixture
(dotted lines) which shows similar peak positions and heights to the ones experimentally measured (solid lines). The sequence position dependency
of the prediction error is shown in Figure \ref{fig:MixChromatogram}.B. On the region of bins 125-700 the prediction
shows high accuracy, with an average root square error of 0.08.  The loss of accuracy at longer sequence positions stems from a cumulative drift
in predicted peak positions, as well as reduced measurement accuracy. We therefore used the region of bins 125-700 for the \bcs reconstruction.

Results of the reconstruction are shown in Figure \ref{fig:MixChromatogram}.C. The algorithm successfully identifies three of the five bacteria
(\textit{Vibrio fischeri, Enterococcus faecalis} and \textit{Photobacterium leiognathi}). Out of the two remaining strains, one (\textit{Staphylococcus epidermidis})
is identified at the genus level, and the other (\textit{Escherichia coli}) is mistakenly identified as \textit{Salmonella enterica}. While \textit{Escherichia coli}
and \textit{Salmonella enterica} show a sequence difference in 33 bases over the PCR amplified region, only two bases are different in the region used for
the \bcs reconstruction, and thus the \textit{Escherichia coli} sequence was removed in the database preprocessing stage. When this sequence is manually
added to the database (in addition to the \textit{Salmonella enterica} sequence), the \bcs algorithm correctly identifies the presence of \textit{Escherichia coli}
rather than \textit{Salmonella enterica} in the mixture. Another strain identified in the reconstruction - the Kennedy Space Center clone KSC6-79 - is highly
similar in sequence (differs in five bases over the region tested) to the sequence of \textit{Staphylococcus epidermidis} used in the mixture.

\section{Discussion}
In this work we have proposed a framework for identifying and quantifying the presence of bacterial species in a given population using
information from a single Sanger sequencing reaction. We have studied the amount
of information present in a current database of the \rRNA gene sequences and the ability of using this information for unique
reconstruction. Essentially, the amount of information needed for identifying the species present in the mixture is logarithmic
in the database size \cite{Candes:01,Donoho:01}, as long as the number of the species present in the mixture is kept constant.
Therefore, a single sequencing reaction with hundreds of bases contains in principle a very large amount of information and
should suffice for unique reconstruction even when the database contains millions of different sequences. Compressed Sensing
enables the use of such information redundancy through the use of linear mixtures of the sample. However, the mixtures need
to be RIP in order to enable an optimal extraction of the information. In our case, the mixtures are dictated by the sequences in the
database, which are clearly dependent. While two sequences which differ in a few nucleotides clearly do not contribute to RIP,
even a single nucleotide insertion or deletion completely shuffles the mixture matrix, thus enabling more efficient reconstruction
through \cs\!\!. Simulation results with noise levels comparable to the measured noise in real chromatograms indicate that this
method can reconstruct mixtures of tens of species. When not enough information is present in the sequence (for example when
a large number of sequences in present in the mix), the reconstruction algorithm's performance  decays gracefully, and still
retains detection of the prominent species.

An important challenge we have encountered when implementing our method on experimental DNA mixtures is in the preprocessing of the
Sanger sequence chromatogram. Both amplitude and position of the peaks are local-sequence dependent,
and therefore corrections are needed in order to attain correct conversion of the raw chromatogram to the mixed-sequence data
used as the input for the reconstruction algorithm. While the effect of local-sequence context on peak {\it height} can be easily incorporated into
the \bcs framework, the effect on peak {\it position} is more complicated to overcome. Since shifting of peak positions is independent for each
sequence present in the mixture, this may result in an cumulative loss of phase in the peaks of the mixed sequence, thus preventing the calculation
of the correct PSSM.
We have overcome this problem by preprocessing the mixed chromatogram data (and the sequence database) at constant sized bins rather than
peak-dependent nucleotide positions. In the current implementation, this preprocessing has enabled us to use approximately 600 nucleotides out of
the experimental mixed chromatogram. The cumulative drift in peak position prediction is currently the limiting factor for performance of
the \bcs algorithm on experimental mixtures, as a difference of even one base position leads to a total shuffling of the PSSMs.

Since the reconstruction problem is mapped to an ordinary \cs notation, generic \cs tools can be applied.
One problem we have faced when using the GPSR algorithm is the large memory needed for handling large
problems, which forced us to remove closely similar sequences in the 16S rRNA database preprocessing step, in order to reduce the problem's size.
Developing and improving \cs solvers is a highly active field of research, and alternative more efficient
\cs solvers such as the greedy matching pursuit approach \cite{tropp2004greed} might enable tackling larger problems, without the removal of similar sequences.
This has the potential to improve reconstruction results, as we have demonstrated when considering the
E. Coli example, which was removed in the database preprocessing, but recovered correctly when considered as input to the \cs algorithm.
In the current implementation, the fact that bacteria have only non-negative frequencies
was not explicitly enforced in eq. (\ref{eq:GPSR}). Utilizing this information is known to simplify the reconstruction
problem \cite{Donoho:positive}, and we expect it to lead to more efficient and accurate reconstruction.
In the current application the mixing matrix does not fulfill the desired RIP condition as it is predetermined by the sequence database, and therefore
has a high coherence which was shown to reduce performance compared to using random sequences. This coherence issue may be
addressed by using novel techniques of dictionary preconditioning \cite{schnass_dictionary_2008}, which improvee sparse signal representations
in redundant dictionaries.

The proposed method can easily be extended to more than one sequencing reaction per mixture, by simply
joining all sequencing results as linear constraints. Such extension can lead to a larger number of linear constraints,
and thus increasing accuracy in cases such as when deciphering larger mixtures in the presence of experimental noise.
For example, using an additional mixed Sanger sequencing run on the sample mixture with the 8F universal primer
(instead of the 1510R primer) could enable to more easily differentiate between the \textit{E. coli} and \textit{S. enterica} strains,
which differ mainly in the beginning of the 16S rRNA sequence. Additionally, combination of several sequencing runs can
enable reconstruction of sequences using different universal primers, thus enabling detection of multiple
strains not amplified by a single universal primer. Usage of additional primers for sequencing can contribute
information even when the sequenced regions overlap, since different sequences are aligned with each
sequencing primer, thus creating a different shuffling of the constraints.
When using only the 16S rRNA gene we cannot differentiate between species with identical 16S genes. For example, it was shown that identification via 16S rRNA
sequencing enabled correct identification (up to the species level) of 95\% of 328 clinical isolates \cite{LeslieHall04012003}, with the remaining 5\% having
non-unique sequences. In order to achieve more accurate identification of such closely related species (or sub-populations  within a given species),
data obtained from sequencing several genes, such as Multi-Locus Sequence Typing (MLST) \cite{maiden1998multilocus}, can also be used in the \bcs framework.
This approach can also be used when many short and inaccurate sequence fragments, which do not suffice for
unique identification of each strain, are present (such as in the case of next generation sequencing methods
\cite{huse_exploring_2008}). Using a sequenced region larger than a single read length and combining the linear constraints
for all the short sequences present may enable more accurate reconstruction using sparseness as the goal function.

While limited to the identification of species with known \rRNA sequences, the \bcs approach may enable low cost simple
comparative studies of bacterial population composition in a large number of samples.

\section*{Acknowledgments}
We thank Amit Singer, Yonina Eldar, Gidi Lazovski and Noam Shental for useful discussions, Eytan Domany for critical
reading of the manuscript, Joel Stavans for supporting this research and Chaime Priluski for assistance with
chromatogram peak prediction data.

% The bibtex filename
\bibliography{bib_metagenomics}

\pagebreak

\begin{flushleft}
{\Large
\textbf{Appendix}
}
\end{flushleft}

\begin{appendix}

\newcounter{append_ctr}

\section {Chromatogram Preprocessing}
\begin{figure}[!ht]
\begin{center}
\includegraphics[totalheight=0.7\textheight]{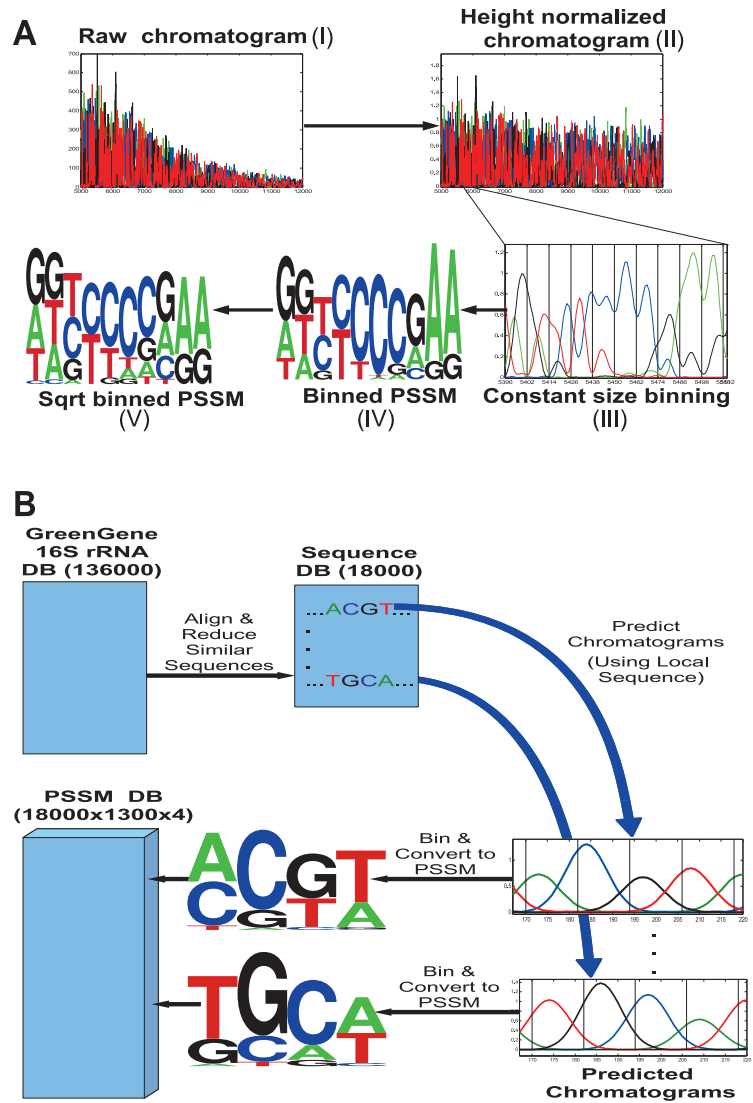}
\end{center}
\caption{
\textbf{Preprocessing steps}
\textbf{A.} Preprocessing of the experimental chromatogram. The result of the Sanger-sequencing of a bacterial mixture (I) is normalized by division
with a $\sim\! 1000$ pixel total intensity running average to compensate for the peak amplitude decrease. The resulting chromatogram (II) is binned into
constant sized bins (sample section shown in III), and the resulting PSSM (sample section shown in IV) is further square-root transformed to obtain
the final experimental PSSM (sample section shown in V).
\textbf{B.} Preprocessing of the 16S rRNA sequence database. Sequences are first aligned and similar sequences are removed. Then, a predicted
chromatogram is generated for each sequence in the database, based on local sequence statistics collected from a training set. Finally, the predicted chromatograms
are binned into constant sized binned and the resulting PSSMs are further square-root transformed in similar to (A), to produce the final PSSMs
which are stored in the database.
}
\label{fig:PreprocessingScheme}
\end{figure}

In order to apply the \bcs algorithm on the experimentally measured mixture chromatogram, several preprocessing steps are required. The purpose
of the chromatogram preprocessing step is to convert the measured chromatogram to a PSSM representing
the frequency of each base at each position in the mixture (see Figure \ref{fig:PreprocessingScheme}.A).

The input to the chromatogram preprocessing is the measured chromatogram, consisting of four fluorescent trace vectors
$\mathfrak{a}, \mathfrak{c},\mathfrak{g}, \mathfrak{t}$, where for example $\mathfrak{a}_p$ represents the signal intensity
for nucleotide $'A'$ at the $p$'s position along the chromatogram, where each position is represented by one pixel in the chromatogram image.
The value $p$ corresponds roughly to the timing of the sequencing reaction, with a resolution of approximately a dozen points per nucleotide,
thus $p$ runs from $1$ to $\sim\! 12 k$.

In a typical Sanger sequencing reaction, the chromatogram peak heights decrease at higher $p$ values (nucleotides further in the sequence which were sequenced
later in the sequencing reaction) due to depletion of the dideoxynucleotides. To overcome this long-scale decrease in signal amplitude, prior to the binning step,
the amplitude at each position was normalized by division with average total peak height in a $\sim 50$ base-pair (bp) region around each position (see step $1$ in the algorithm
description below).

The resulting vectors after the normalization step are binned into constant sized bins, and the sum of intensity values of each bin is computed for the four different
nucleotides. Then, we take square root of this sum for the four different nucleotides for the $i$'th bin as the $i$'th column in the output $4 \times k$ PSSM.  The square root
is used rather than the sum as this was shown to decrease the effect of large outliers. The resulting $4 \times k$ PSSM is used as input to the \bcs reconstruction.
Formally, the preprocessing algorithm is described below:

\nin {\bf Algorithm:} Chromatogram Preprocessing \\
\nin {\bf Input:} $(\mathfrak{a}, \mathfrak{c},\mathfrak{g}, \mathfrak{t})$ - four fluorescent trace vectors \\
\nin {\bf Output:} $P = (a, c, g, t)^t$ - a PSSM representing nucleotide frequencies
\begin{enumerate}

\item Normalize the chromatogram amplitude:
\be
\mathfrak{a}_p = \frac{50 \cdot 12 \cdot \mathfrak{a}_p}{\sum_{q=-25 \cdot 12}^{25 \cdot 12} (\mathfrak{a}_{p+q} + \mathfrak{c}_{p+q} +
\mathfrak{g}_{p+q} + \mathfrak{t}_{p+q})}
\label{eq:normalization_seq_position}
\ee
and similarly for $\mathfrak{c}_p$,$\mathfrak{g}_p$ and $\mathfrak{t}_p$.

\item bin into constant sized bins, average bin values and apply square root transformation:
\be
a_i = \sqrt{\sum_{p=12i}^{12i+11}\mathfrak{a}_p} \:, \quad i=1...k
\label{eq:sqrt_binning}
\ee
and similarly for $c_i$,$g_i$,$t_i$.

\end{enumerate}

\clearpage
\newpage

\section{Database Preprocessing}
The purpose of the Database Preprocessing scheme is to produce predicted PSSMs for all 16S rRNA sequences in the databsase (see Figure \ref{fig:PreprocessingScheme}.B).
These predicted PSSMs are used later in the reconstruction algorithm as 'basis vectors' whose linear combination with the appropriate
coefficients gives the PSSM obtained from observed mixed chromatogram, as described in the previous section.
An important intermidiate stage in obtaining the predicted PSSMs is generation of predicted chromatograms for the database sequences.
When generating predicted chromatograms we take into account local sequence context - the estimation of this sequence context effects requires a training set of real chromatograms from known sequences from which chromatogram peak height and positions statistics are computed as a preliminary step.
The database preprocessing therefore contains two steps:
\begin{enumerate}
\item A preliminary step: Compute local-sequence adjusted chromatogram statistics. \\
Here we use a training set $S'$ of known sequences and their chromatograms as input and compute tables $H$ and $P$ representing chromatogram
peak heights and positions, respectively, for different local-sequence contexts.

\item Database PSSMs Generation step: Generate a database of predicted PSSMs. \\
Here the input is a set $S$ of $N$ sequences and the pre-computed tables $H$ and $P$. The sequence input and tables are used
to determine peak heights and positions and compute a set of $N$ chromatograms of the form $(\mathfrak{a}, \mathfrak{c},\mathfrak{g}, \mathfrak{t})$,
one for each sequence in the database. These chromatograms are then further processed to get predicted PSSMs.
This step is illustrated in Figure \ref{fig:PreprocessingScheme}.B.
\end{enumerate}

After performing these two steps, and additional alignment step is required to match the start
of the measure and predicted chromatograms.

\subsection{Preliminary Step: Compute local-sequence adjusted chromatogram statistics}
In the course of the Sanger sequencing process, both the polymerase specificity for incorporating deoxynucleotides over
dideoxynucleotides and the fragment mobility depend on sequence local to the incorporation point. Therefore for each nucleotide in the
DNA fragment being sequenced, its corresponding chromatogram peak height $a$ and position $b$ are affected by the preceding nucleotides
\cite{lipshutz_dna_1994}. In order to predict and correct for the effect of local sequence context on the resulting chromatogram,
we collected statistics from a training set $S'$ of $1000$ sequencing runs performed on an ABI3730 machine. Runs were obtained from various
Sanger sequencing experiments performed by different labs and for different organisms thus presenting us with a diverse genomic
training set. The average length of the runs was approximately $800$ base-pairs. Chromatogram heights were normalized to overcome the long scale
amplitude decrease (as described in the Chromatogram Preprocessing section) and chromatogram peaks were identified. We use $p_{i,j}$ and $h_{i,j}$
to denote the position and height of the $j$-th nucleotide in the $i$-th sequence, respectively (we have chosen the top peak out of the four traces representing
different bases - since each chromatogram represented a single sequence, these peaks were in most cases clearly higher and distinguishable from the three
other peaks).

We have modeled the local sequence context by looking at the
$5$ nucleotides preceding each nucleotide, giving us $4^6 = 4096$ different unique 6-mers, each representing a possible nucleotide and the $5$ nucleotides
preceding it. For each unique 6-mer, we have searched for all of its occurrences in the $1000$ sequences, and averaged the peak height and position data
of the last nucleotide over all such occurrences in the $1000$ sequences analyzed. We have used 6-mers as this gives the maximal
context length for which we had sufficient statistics to collect for each bin (approximately $200$ instances per bin, on average) - it is possible that smaller
context is sufficent for accurate prediction of chromatogram heights.
More formally, for a given kmer $\alpha = (\alpha_1,..,\alpha_6)$, we have computed $H(\alpha)$ and $P(\alpha)$ as follows.
The peak height statistic $H(\alpha)$ measures the corresponding local-averaged peak heights:
\be
H(\alpha) = \frac{\sum_{i,j}  1_{\{\alpha_1 = S'_{i,j-5}, ..,\alpha_6 = S'_{i,j}\}} h_{i,j}}  {\sum_{i,j}  1_{\{\alpha_1 = S'_{i,j-5}, ..,\alpha_6 = S'_{i,j}\}}}
\label{eq:peak_height}
\ee

\nin For peak position $P(\alpha)$, we first computed the relative peak-peak distance for each position:
\be
d_{i,j} = \frac{p_{i,j}-p_{i,j-1}}{\sum_{j=2}^k p_{i,j}-p_{i,j-1}}.
\ee

\nin Then, we measured the average relative peak-peak distance between the current peak and the previous peak:
\be
P(\alpha) = \frac{\sum_{i,j}  1_{\{\alpha_1 = S'_{i,j-5}, ..,\alpha_6 = S'_{i,j}\}} d_{i,j}}   {\sum_{i,j}  1_{\{\alpha_1 = S'_{i,j-5}, ..,\alpha_6 = S'_{i,j}\}}}
\label{eq:peak_position}
\ee

\begin{figure}[!ht]
\begin{center}
\includegraphics[totalheight=0.25\textheight]{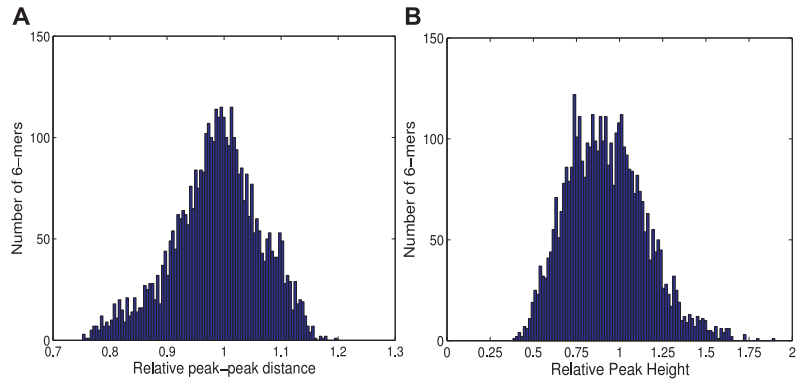}
\end{center}
\caption{
\textbf{Local-sequence effect on chromatogram peak height and position.}
\textbf{A.} Distribution of average normalized peak-peak distances for the 4096 sequence 6-mers. % ??? ADD DETAILS
\textbf{B.} Distribution of normalized peak heights for the 4096 sequence 6-mers. Both distributions show a rather wide spread around one,
showing that local sequence context has a significant effect on peak height and position.}
\label{fig:PosHeightDist}
\end{figure}

The results were the final peak height and position tables $H$ and $P$ respectively, each of size $4096 (=\!4^6)$ (available on the article website).
While the average height and position were $1$ (as was ensured by our normalizations), there was significant variability in height and position
according to sequence context, with height values $H$ typically in the range $\sim\! 0.5\!-\!1.3$ and position values $P$ in the range $\sim\! 0.8\!-\!1.2$
(see Figure \ref{fig:PosHeightDist}).
An additional sequence-independent non-linearity in the peak position was observed in the chromatograms studied, where distance between consecutive peak increases
as we move further along the chromatogram. This was accounted for by fitting an additional linear model based only
on the nucleotide sequence position, giving an additional parameter of $\beta = 0.00036$ representing increase in peak-peak distance
with each position (see next section in eq. (\ref{eq:peak_params_estimation}) ).

\clearpage
\newpage

\subsection{Database PSSMs Generation step: Generate a database of predicted PSSMs.}
The input for this step is the database sequence matrix $S$ and the output is a set of PSSMs $(a, c, g, t)^t$,
one for each sequence $S_i$ in the database. The database processing scheme is applied only once to the database and the predicted PSSMs are
stored and can be used for any new mixture sample obtained. It is applied  to each sequence in the database independently.

For every nucleotide in the database, we estimated it's chromatogram peak height and peak-peak distance as:
\be
a_{i,j}=H( L_{i,j} ), \quad b_{i,j}=b_{i,j-1}+P (L_{i,j} )+ \beta j
\label{eq:peak_params_estimation}
\ee
where $L_{i,j} = (S_{i,j-5},..,S_{i,j})$ denotes the local 6-mer sequence context of nucleotide $j$ in the $i$-th sequence ($L_{i,j} \in 1 ... 4^6 $).
The parameter $\beta = 0.00036$ represents the increase in peak-peak distance per position as we move further along the sequence.

In order to generate a chromatogram trace for a given sequence, each peak was modeled as a Gaussian centered at the peak position and with height equal to the peak height.
Thus, for every nucleotide in a sequence, a corresponding peak was created in the chromatogram using the Gaussian peak function
\be
f_{i,j}(x)=a_{i,j} e^{- \frac{(x-b_{i,j})^2}{2c^2}}
\label{eq:gaussian_peak}
\ee

The widths of the chromatogram Gaussian peaks were approximated by using a constant peak width obtained by setting $c=0.4$.
A chromatogram was generated for each sequence by summing the values of obtained  $f_{i,j}$ over all nucleotides. Each $f_{i,j}$
was evaluated for $x$ values equally spaced in the range $[0,k]$, at a resolution of $1/12$ thus giving $12 k$ different $x$ values $x_1,..,x_{12 k}$.
Each  $f_{i,j}$ contributed to the sum only for the trace of the corresponding base in the sequence, and for the other three bases the contribution to the sum was zero.
That is, the trace vector for the nucleotide 'A' for the $i$-th sequence was computed as:

\be
\mathfrak{a}_{p} = \sum_j f_{i, j } (x_p) 1_{\{S_{i,j} = 'A'\}}
\ee

and similarly for the other three nucleotides.
The resulting predicted chromatograms were binned using a constant bin size of $1$ and transformed via square root, in similar
to the chromatogram preprocessing step in eq. (\ref{eq:sqrt_binning}), to give a PSSM database with one PSSM for each sequence in $S$.

\subsection{Alignment of Predicted and Measured Chromatograms}
\begin{figure}[!ht]
\begin{center}
\includegraphics[totalheight=0.22\textheight]{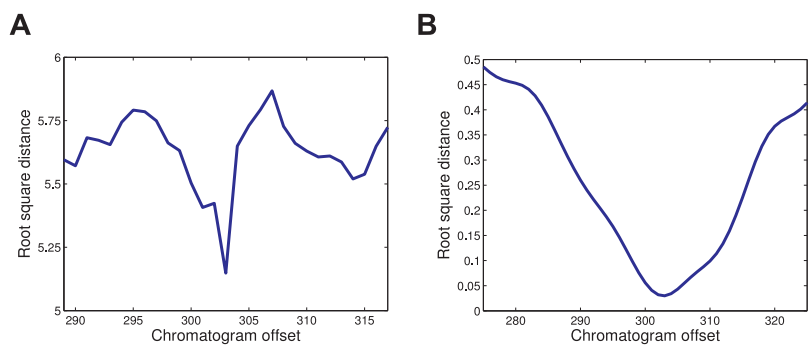}
\end{center}
\caption{
\textbf{Determination of chromatogram offset.}
\textbf{A.} Root square distance between measured chromatogram and the chromatogram predicted from the \bcs reconstruction.
Minimal value is obtained when position 1 in the measured chromatogram is aligned to position 304 in the database.
\textbf{B.} Root square distance between measured chromatogram and the chromatogram predicted using the known composition
of the five species in the mixture. Minimal value is obtained when position 1 in the measured chromatogram is aligned to position 304 in the database.
}
\label{fig:ChromatogramOffset}
\end{figure}

Sanger-sequencing chromatograms display an initial region ($\sim\!\! 100$ bases) which is highly noisy and therefore unusable.
We are therefore faced with the problem of correctly aligning the initial bin position in the measured chromatogram and
the bin positions of the predicted chromatograms.
This was solved by trying the \bcs reconstruction for different initial bin offsets in the measured chromatogram,
and selecting for the offset with the lowest reconstruction root square distance (see Figure \ref{fig:ChromatogramOffset}.A).
This reconstruction root square distance is calculated as the difference between the measured chromatogram and the
predicted chromatogram based the reconstructed species frequencies. To verify the validity of this criterion, we also
compared the average distance between the measured chromatogram and the predicted mixture chromatogram obtained using
the known mixture composition (see Figure \ref{fig:ChromatogramOffset}.B), using various offsets for the measured
chromatogram binning. Both methods obtained an identical offset, which was used in the reconstruction.

\clearpage
\newpage

\begin{figure}[!ht]
\begin{center}
\includegraphics[totalheight=0.5\textheight]{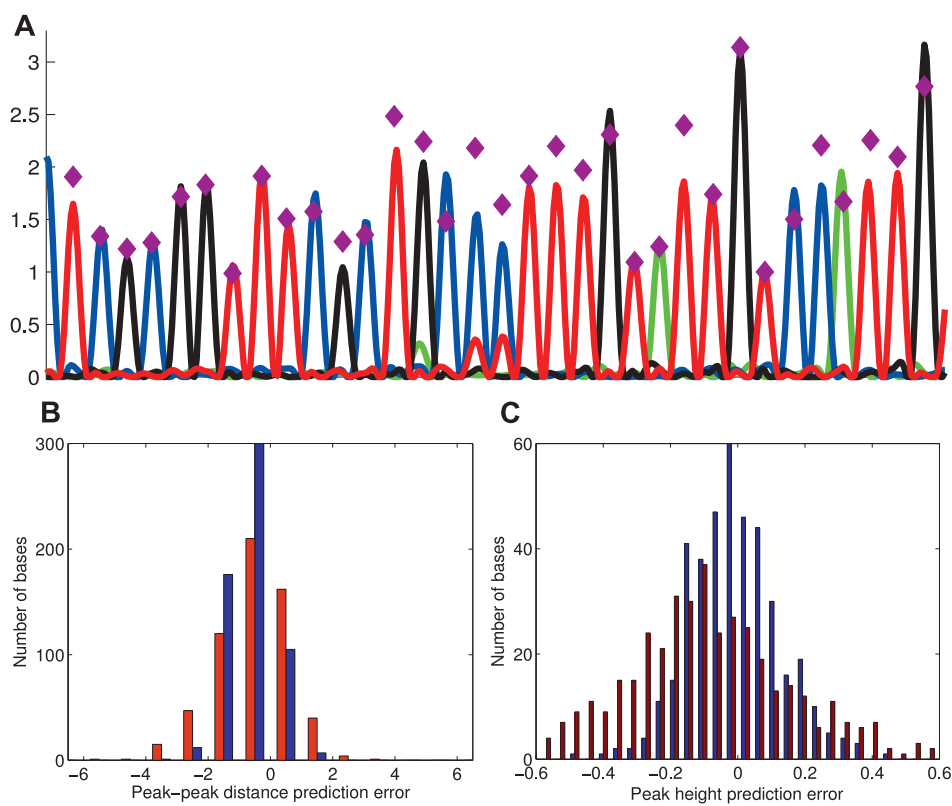}
\end{center}
\caption{
\textbf{Effect of local sequence on chromatogram peak heights and positions.}
\textbf{A.} Sample sequenced chromatogram and prediction (magenta circles) of peak heights and positions based on local (6-mer) sequence.
\textbf{B.} Distribution of peak-peak distance differences between predicted and measured peak positions before (red) and after (blue) correction for local
sequence effects. The average peak-peak distance is $\sim 12$ pixels.
\textbf{C.} Distribution of distance between predicted and measured peak heights before (red) and after (blue) correction for local
sequence effects. Employing local sequence context improves both height and positions predictions.}
\label{fig:SampleChromatogram}
\end{figure}

\clearpage
\newpage

\end{appendix}

\end{document}